\documentclass[twocolumn,secnumarabic,amssymb, nobibnotes, aps, prd]{revtex4-1}

\newcommand{\mycomment}[1]{}

\usepackage{amsmath, braket, xcolor, ulem, graphicx}

\begin{document}
\title{Laser mode-hopping assisted all-optical single beam pulsed atomic magnetometer }
\author{Ji Hoon Yoon, Sang Hyuk Hong, Taek Jeong, Sin Hyuk Yim, Kyu Min Shim, and Sangkyung Lee}
\email{sklee82@add.re.kr}
\affiliation{Emerging Science and Technology Directorate, Agency for Defense Development, Republic of Korea}

\begin{abstract}
We demonstrate an all-optical single beam pulsed atomic magnetometer assisted by laser mode-hopping in a distributed Bragg reflector (DBR) laser. We implement a temporal sequence of the laser current, with sinusoidal current modulation including the laser mode-hop current for synchronous optical pumping, and a following constant current for paramagnetic Faraday rotation measurements, to probe the free induction decay (FID) of transverse $^{87}$Rb spin polarization. Repetitive sudden frequency shifts of 20 GHz around the pressure-broadened $^{87}$Rb spectra, originating from laser mode-hopping, enable discontinuous optical pumping modulation with a large depth, which enhances transverse spin polarization. We achieved a sensitivity of 0.6 pT/Hz$^{1/2}$ in a magnetic field of 27 $\mu$T, mainly limited by the photon-shot-noise and the magnetic field noise induced by the current noise in the current supply for driving the bias magnetic field coil. The Cramer-Rao lower bound (CRLB) of the sensitivity due to the non-magnetic noise such as photon shot-noise is 131 fT/Hz$^{1/2}$. Our approach based on laser mode-hopping can be applied for the miniaturization of all-optical atomic magnetometers with sub-pT/Hz$^{1/2}$ sensitivities. 
\end{abstract}
\maketitle

\section{Introduction}
Laser mode-hopping refers to a jump in lasing mode which stems from a relative frequency shift in longitudinal cavity modes to the gain curve. It originates from changes in temperature or injection current and is generally considered to be problematic in laser spectroscopy and telecommunications because it is accompanied by unwanted intensity fluctuations and large sudden frequency shifts. Many studies have attempted to extend the mode-hop free range to facilitate continuous wavelength tuning \cite{gong2014mode, dutta2012mode, fuhrer2008extension}. There have also been attempts to study the dynamics of laser mode-hopping to better understand the underlying physics of mode-hopping.  \cite{kundu2018ultrafast, butler2017direct, happach2020effect}. 

A large sudden frequency shift feature during laser mode-hopping, however,  may be useful in some applications. For example, in Bell-Bloom type atomic magnetometer experiments where the modulation of optical pumping rate to induce spin-precession is involved, large sudden frequency shifts during mode-hopping can enhance spin-polarization by increasing the modulation depth of the optical pumping. In order to build up large transverse spin polarization in all-optical Bell-Bloom type atomic magnetometers, a variety of techniques have been explored, including sinusoidal modulation of the laser frequency  \cite{kimball2017situ, jimenez2009sensitivity}. Some advancements in the development of the nonlinear magneto-optical rotation (NMOR) magnetometer have been reported, based on pulse-shaped amplitude modulation of the beam, which takes advantage of the discontinuous modulation \cite{gawlik2006nonlinear, gerginov2017pulsed}. Obtaining a discontinuous lineshape in the modulation function, however, requires cumbersome optics such as acousto-optic modulators (AOM) or intricate techniques to generate a square-wave RF magnetic field \cite{ranjbaran2017effects, hunter2018free}. 

Frequency switching by means of mode-hopping, on the other hand, only requires weak sinusoidal modulation of the laser diode (LD) injection current to create a large discontinuous optical pumping modulation. Its effectiveness can be estimated from the optical pumping rate formula. The optical pumping rate is a function of laser intensity and detuning, given by $R \propto I/(\Delta\nu^2 + (\Gamma_L/2)^2)$ where $I$ is the laser intensity, $\Delta\nu$ is the laser detuning, and $\Gamma_L$ is the line broadening. For example, when the frequency shift during mode-hopping is $2\Gamma_L$, the optical pumping rate of $\Delta\nu = 2\Gamma_L$ is 17 times smaller than that of $\Delta\nu = 0$. To achieve the same modulation depth using intensity modulation, the laser intensity has to be changed from $I$ to $I/17$. To obtain such a large modulation depth, an external light modulator, for example, AOM or electro-optic modulator (EOM), can be applied. To miniaturize the device, current modulation can be directly injected into the LD instead of using an external light modulator. Since the linear frequency shift of the LD is a function of injection current, in our case, -1 GHz/mA in the mode-hop free range, an amplitude of current modulation of 20 mA is required to achieve a 20 GHz detuning. This corresponds to a doubling of  $\Gamma_L = 10$ GHz, which is a typical line broadening of the Rb D$_1$ line in nitrogen buffer gas. Such detuning can be easily obtained with only a few mA when laser mode-hopping is adopted. In addition, the injection current into the LD does not need to be discontinuous. A continuous laser modulation with a small modulation depth leads to a large depth of discontinuous optical pumping modulation when laser mode-hopping is applied. This is an advantage of laser mode-hopping assisted optical pumping, because a large amplitude of discontinuous laser injection current can be a source of unwanted magnetic field noise.

The frequency switching feature can also be helpful for implementing single beam all-optical atomic magnetometer based on FID measurements. By utilizing laser mode-hopping  we can easily shift the laser frequency to be far detuned from, or to be resonant with, the atomic resonance.  A temporal sequence of the laser current containing the mode-hop current enables FID measurements. In the on-resonance period when the laser is resonant with the atomic transition, the atomic spin polarizes along the direction of the laser beam. During the following off-resonance period, when the laser is far detuned from the atomic transition due to the mode-hopping, optical pumping stops and induces the FID of the transverse atomic spin. The far-detuned laser experiences paramagnetic Faraday rotation depending on the transverse atomic spin that precesses with the Larmor frequency. 

Usually, single beam atomic magnetometers utilize a magnetic field modulation technique using a coil \cite{liu2020single,ding2020single,oelsner2022integrated}. Their structure is very simple but they can suffer from magnetic field interference when forming a gradiometer or a magnetometer array. To diminish the effect of magnetic interference, all-optical atomic magnetometers have been widely developed. An all-optical vector atomic magnetometer was reported \cite{patton2014all}. An all-optical atomic magnetometer with a co-linear configuration to measure biomedical magnetic fields was implemented by placing the pump and probe lasers at a slight angle within the same window \cite{limes2020portable}. Single beam all-optical atomic magnetometers based on an external light modulator have also been studied \cite{petrenko2022towards, hunter2018free}. And an all-optical single beam atomic magnetometer based on direct current modulation to the vertical-cavity surface-emitting laser (VCSEL) has already been demonstrated with a sensitivity level of 10 pT/Hz$^{1/2}$ \cite{jimenez2009sensitivity, hunter2018free}. However, the low power of the VCSEL is insufficient and unable to achieve a significant optical pumping rate $R(t)$ so that its sensitivity is limited. 

Our strategy for a single beam all-optical atomic magnetometer without the use of an external light modulator involves adopting a DBR laser instead of a VCSEL to achieve a higher optical pumping rate, and applying a current sequence including laser mode-hopping. It consists of an optical pumping stage and a following probing stage to measure the FID of macroscopic $^{87}$Rb spin.

In this paper, we demonstrate an all-optical single beam pulsed atomic magnetometer utilizing the mode-hopping of a DBR laser. We employ mode-hopping to selectively switch the laser frequency between on-resonance and off-resonance with the \({}^{87}\)Rb D$_1$ line so that a sequence of optical pumping and FID probing is realized. We observe the FID as a function of offset, modulation amplitude of current modulation and analyze the discontinuous optical pumping effect. Additionally, we measure the sensitivity of our magnetometer and compare it to the CRLB considering non-magnetic noise, demonstrating the achievable sensitivity of our magnetometer.

\section{Model for discontinuous optical pumping}

\begin{figure}
\centering
\includegraphics[width=0.98\linewidth]{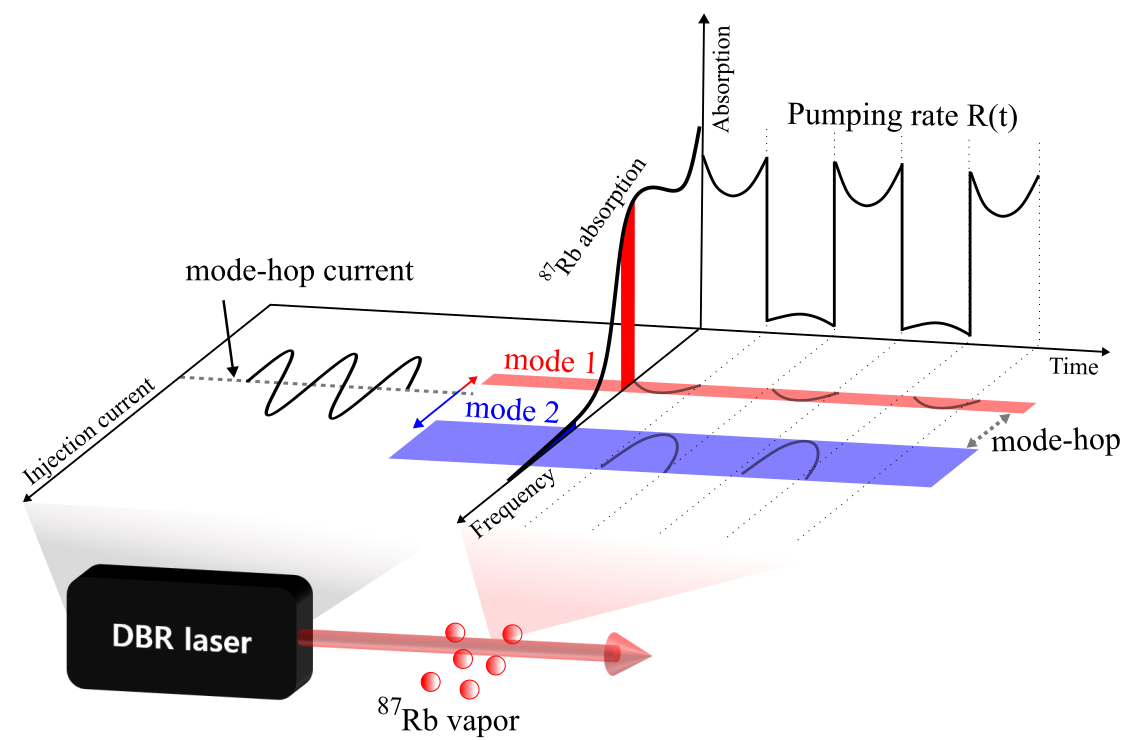}
\caption{Illustration of the concept of frequency modulation assisted by laser mode-hopping. A sinusoidal modulation of laser injection current is converted to a discontinuous output frequency profile. The red and blue areas represent the adjacent cavity modes in a DBR laser. A small variation around the mode-hop current results in a huge shift in laser output frequency. The frequencies of these modes correspond to the resonant and off-resonant conditions of the pressure-broadened ${}^{87}$Rb absorption profile, respectively. The repetitive jumps between cavity modes induces discontinuous synchronous optical pumping.}
\label{fig:schematic} 
\end{figure}

A discontinuous optical pumping profile can be achieved by incorporating a simple sinusoidal modulation of the laser injection current with assistance of mode-hopping (see Figure~\ref{fig:schematic}). When the sinusoidal current modulation contains the mode-hop current, the laser frequency jumps by the frequency difference between cavity modes.  Assuming that the frequency of the mode 1 (or the mode 2) is resonant with (or far-detuned from) the atomic resonance, the optical pumping rate, $R \propto I/(\Delta\nu^2 + (\Gamma_L/2)^2)$,  will exhibit steep changes, as shown in Fig.~\ref{fig:schematic}. The change in optical pumping rate is sufficiently sudden, in other words, discontinuous, because the switching time of mode-hopping, on the order of 10 ns, is much shorter than the Larmor period of 2.85 $\mu$s in a magnetic field of $50~\mu$T \cite{butler2017direct}.

In order to analyze the effect of the discontinuous optical pumping induced by a mode-hopping, we investigate the macroscopic spin Bloch equation including the optical pumping modulation term, $R(t)$. 

The Bloch equation model, widely used for atomic magnetometers, is suitable for modeling arbitrary modulations distorted by mode-hopping \cite{grujic2013atomic}. The Bloch equation is given by

\begin{equation}
\frac{d\langle \vec{S}\rangle}{dt} =\gamma\langle \vec{S}\rangle \times  B_y \hat{y}- \Gamma_{re} \langle \vec{S}\rangle+R(t)(\frac{\vec{s}}{2}-\langle \vec{S}\rangle).
\label{BlochEq}
\end{equation}

\noindent
Here, \(\vec{s}=(e^* \times e)/i\) represents the photon spin vector where $e$ represents a unit polarization vector in the yz plane so that $\vec{s} = s \hat{x}$. \(\Gamma_{re}\) is the spin relaxation rate, mainly contributed by the spin exchange collisions between \({}^{87}\)Rb atoms. \(R(t)\) is the optical pumping rate for unpolarized atoms, depending on the laser detuning from the $D_1$ resonance and the laser intensity \cite{vasilakis2011precision}.  $\gamma$ is the gyromagnetic ratio. We consider the magnetic field to be directed in the \(\hat{y}\) axis.

Any periodic optical pumping rate \(R(t)\) can be expanded by a Fourier series $\sum_{n=-\infty}^{\infty} R_n e^{i n \omega t} $ where $R_n = \frac{1}{T}\int_{-T/2}^{T/2}{dt R(t) e^{i n\omega t}}$ for non-zero $n$. By assuming nearly on-resonant modulation, {\it $\omega \simeq \omega_0 = \gamma|B_y|$ } and assuming that the Larmor frequency $\gamma |B_y|$ is much larger than other characteristic rates such as $\Gamma_{re}$ and $R(t)$, the rotating wave approximation can be applied to the spin Bloch equation. The steady-state amplitude of the rotating transverse spin, \(S_0\), can be written as

\begin{equation}
\label{SpinAmplitude}
S_0 \approx \frac{(R_1+R_{-1})|s|}{R_0} \frac{1}{4(1 +\Gamma_{re}/R_0)}.
\end{equation}

In the case of sinusoidal optical pumping modulation described by \(R(t) = A+ B \cos(\omega t) \), the value of $(R_1+R_{-1})/R_0$ is calculated as $B/A$, which cannot exceed 1 because $R(t)$ is always positive, {\it i.e., $A \geq B$}. The maximum $S_0$ is 1/4 in the sinusoidal optical pumping case when $R_0 = R_1$ and $R_0 \gg \Gamma_{re}$. A discontinuous \(R(t)\) (e.g. square-wave function), on the other hand, allows for \((R_1+R_{-1})/R_0 >1\). 

In the case of a square wave optical pumping modulation, \(R_0\) varies according to the duty cycle \(d\) when the maximum value of \(R(t)\) is fixed. 
Assuming the maximum optical pumping rate of the square wave optical pumping modulation is \(R_m\), then \(S_0\) can be expressed by

\begin{equation}
S_0 \approx  \frac{|s| \text{sinc}(\pi d)}{2(1+\frac{\Gamma_{re}}{ R_m d})}.
\end{equation}

\noindent
The maximum $S_0$ is 1/2 when $d \rightarrow 0$ in the regime where $\Gamma_{re} \ll R_m d$. This implies that a delta-function-like optical pumping induces the largest $S_0$ as long as the peak intensity of the laser pulse is strong enough to satisfy $R_m  \gg \Gamma_{re}/d$. The reason why the delta-function-like optical pumping is optimal can be explained in terms of $\vec{s}$ and $\langle \vec{S}\rangle$. In the Bell-Bloom configuration, optical pumping not only creates the Rb spin polarization along the direction of the pumping beam,but also relaxes the transverse spin. When the direction of $\vec{s}$ does not coincide with $\langle \vec{S}\rangle$, optical pumping destroys the transverse spin precession and aligns the atomic spin with the direction of the pumping beam. It ceases to serve as a transverse spin relaxation mechanism only if $\langle \vec{S}\rangle$  is oriented in the same direction of $\vec{s}$. As a result, we can conclude that the delta-function-like optical pumping is optimal. 

In the regime where the optical pumping rate $R_m$ is not much larger than $\Gamma_{re}/d$, the nonzero $d$ can be optimal. For example, in the weak pumping condition where $\Gamma_{re} = 2\pi \times 685$ Hz, $R_m = 2\pi \times 300$ Hz, and $|s| = 1/2$, the optimal duty cycle is calculated to be around 50 \%. In this case, $S_0$ of the discontinuous optical pumping with a 50 \% duty cycle is $4/\pi$ times larger than that of the sinusoidal optical pumping modulation if the maximum optical pumping rates are the same.

\section{Experimental Setup}

\begin{figure}[t]
\centering
\includegraphics[width=0.99\linewidth]{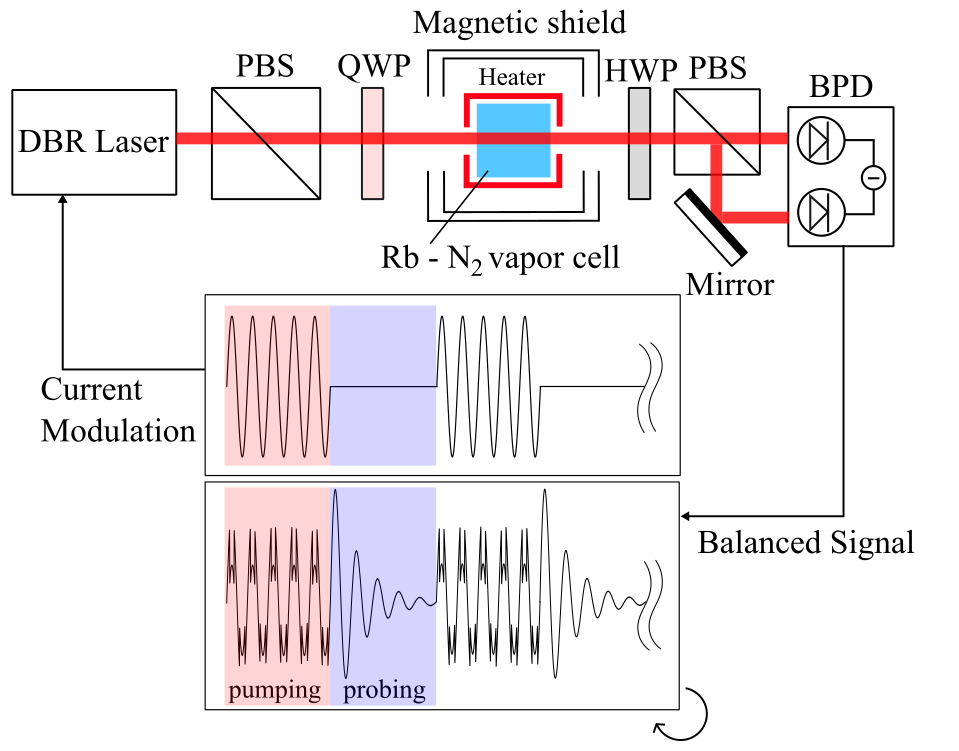}
\caption{Experimental setup for the single beam all-optical magnetometer. DBR laser: Distributed Bragg Reflector laser, PBS: Polarization Beam Splitter, QWP: Quarter Wave Plate, HWP: Half Wave Plate, BPD: Balanced Photodetector. A burst sine wave modulation to the DBR laser realizes pulsed operation at 1 kHz. The red and blue regions show the pumping and probing stages of the pulsed magnetometer, respectively.}
\label{fig:setup} 
\end{figure}

\begin{figure*}
\centering
\includegraphics[width=0.9\linewidth]{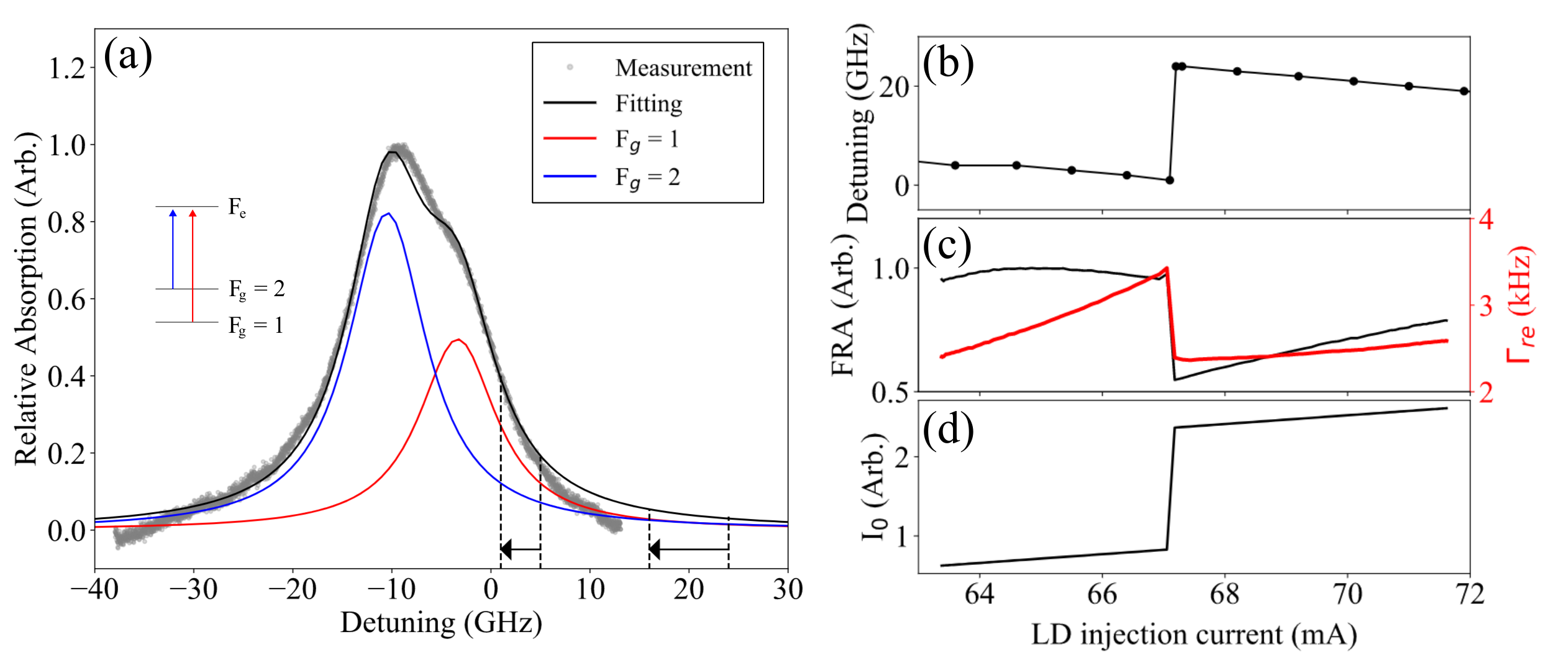}
\caption{(a) The $^{87}$Rb D$_1$ absorption spectra in the atomic vapor cell. The grey dots indicate the absorption of the linearly polarized laser beam, while the black solid line depicts the fitting curve. The Full Width at Half Maximum linewidth $\Gamma_L$ is 10.5 GHz which implies that the pressure of $\textrm{N}_2$ in our cell is about 580 Torr. The red and blue solid lines depict the absorption spectra of hyperfine transitions from $F_g = 1$ and $F_g = 2$, respectively. The black arrows show how the laser frequency detuning changes as the laser current increases. (b) Laser detuning as a function of LD injection currents. The mode-hop current is 67.1 mA. (c) Relative Faraday rotation angle (black solid line) and transverse spin relaxation rate (red solid line) $\Gamma_{re}$ as a function of LD injection current. The injection current below the mode-hop current induces a larger Faraday rotation angle and transverse spin relaxation rate because the corresponding laser frequency is within the $^{87}$Rb D$_1$ absorption spectrum. (d) Relative optical power of the laser beam after passing through the atomic vapor cell as a function of LD injection current. The laser injection current above the mode-hop current gives higher beam power because the off-resonant light experiences weak absorption.}
\label{fig:absorption} 
\end{figure*}

The experimental setup is shown in Figure~\ref{fig:setup}. A DBR laser (Photodigm 794.978 nm series with TOSA package) is used as a light source. Mode-hopping in the DBR laser occurs under fixed conditions due to its monolithic nature, making the jump to the output frequency controllable. Two main factors influence the frequency tuning of the DBR laser output: the temperature of the LD heat sink and the laser injection current. When modulating the  injection current with a stabilized laser temperature, a slight alteration in laser current triggers a transition between two adjacent cavity modes, as illustrated in Fig.~\ref{fig:schematic}. The transition of the lasing mode near the mode-hop current results in a significant jump in the laser output frequency, which was almost 20 GHz in our case.  With the aid of mode-hopping, a sinusoidal modulation of the LD injection current is converted to a discontinuous laser output frequency profile without any additional optics.

The linearly polarized laser beam is first transformed into an elliptically polarized beam after passing through a QWP. The optimal QWP angle to increase the signal is 25 ${}^\circ$ and the corresponding $|s|$ is about 0.67 (see the appendix). The laser beam serves both as the probing and pumping beam. It illuminates a $5 \times 5 \times 5$ m$^3$ cubic cell which contains \(^{87}\)Rb vapor and $\mathrm{N_2}$ gas at a pressure of 580 Torr (see Figure~\ref{fig:absorption}~(a)). 

The homemade atomic gas cell is heated up to 103 $^\circ$C using a two-layer structured pair of AC heaters designed to negate magnetic field noise \cite{yim2018note}. After passing through the atomic gas cell, which is inside a two layered \(\mu\)-metal shield, the laser beam goes through a HWP. We note that a four layered \(\mu\)-metal shield (MS-1, Twinleaf) is applied for sensitivity estimation to further reduce unwanted environment magnetic field noise. A polarimeter, consisting of a PBS and a BPD, measures the degree of polarization rotation of the elliptically polarized beam induced by the paramagnetic Faraday rotation effect. The rotation angle of the polarization is proportional to $\Delta\nu/[\Delta\nu^2 + (\Gamma_L/2)^2]$ where $\Delta\nu$ is the laser detuning in the probing stage.

Our pulsed magnetometer alternates between pumping and probing modes at a repetition rate of 1 kHz. The laser injection current is modulated by a burst sine wave generated by a function generator (SDG2082X, Siglent), realizing a pulsed magnetometer operation. In the pumping stage, depicted as the red region in Fig.~\ref{fig:setup}, the LD injection current is modulated at the Larmor frequency of \(^{87}Rb\) (which is 100 kHz in our case), and incorporates a mode-hop current into the modulation of the LD injection current. The mode-hop current is 67.1 mA when the TEC temperature of the LD is set to 32.05 $^\circ$C. When the LD injection current is above the mode-hop current, the laser output frequency deviates from the Rb resonance line as the laser detuning becomes almost 20 GHz, shown as an arrow in Fig. 3. On the other hand, when the LD current is below the mode-hop current, the laser output frequency gets close to the $F_g = 1$ transition line as the laser detuning becomes almost 0 GHz. Due to the pressure broadening of 10.5 GHz, the laser beam also weakly induces the transitions from $F_g = 2$.

In the probing stage, illustrated by the blue region in Fig. 2, the sinusoidal modulation of the LD injection current ceases and the injection current is fixed at a certain value at which the laser frequency is off-resonant. The detuned laser beam undergoes the paramagnetic Faraday rotation effect where the polarization rotation angle is proportional to the atomic spin projected in the direction of the laser beam propagation.  The laser beam with rotated polarization is directed into a polarimeter.  A damped sine signal on which the FID is imprinted can be obtained from the BPD. It can be described by

\begin{equation}
\label{fittingEq}
V(t, \Delta\nu) = V_0(\Delta\nu) e^{-\Gamma_{re} t} \cos(\omega_0 t + \phi)
\end{equation}

\noindent
where $V_0(\Delta\nu)$ is the FID amplitude which is proportional to $S_0$ multiplied by beam intensity after absorption $I_0$ and depends on the laser detuning $\Delta\nu$ at the probing stage. $\Gamma_{re}$ is the measured spin relaxation rate of the FID signal and $\phi$ is the phase of the FID signal. By fitting the FID signal with Eq.~\eqref{fittingEq}, we can extract the initial amplitude of the FID signal, the spin relaxation rate, and the Larmor frequency. 

In order to estimate the sensitivity of our magnetometer, we apply a frequency counter (53230A, Keysight) which measures the frequency of the signal based on the zero-crossing detection method. The reciprocal of the time between two zero-crossing points becomes the Larmor frequency.  

\begin{figure*}[t]
\centering
\includegraphics[width=0.85\linewidth]{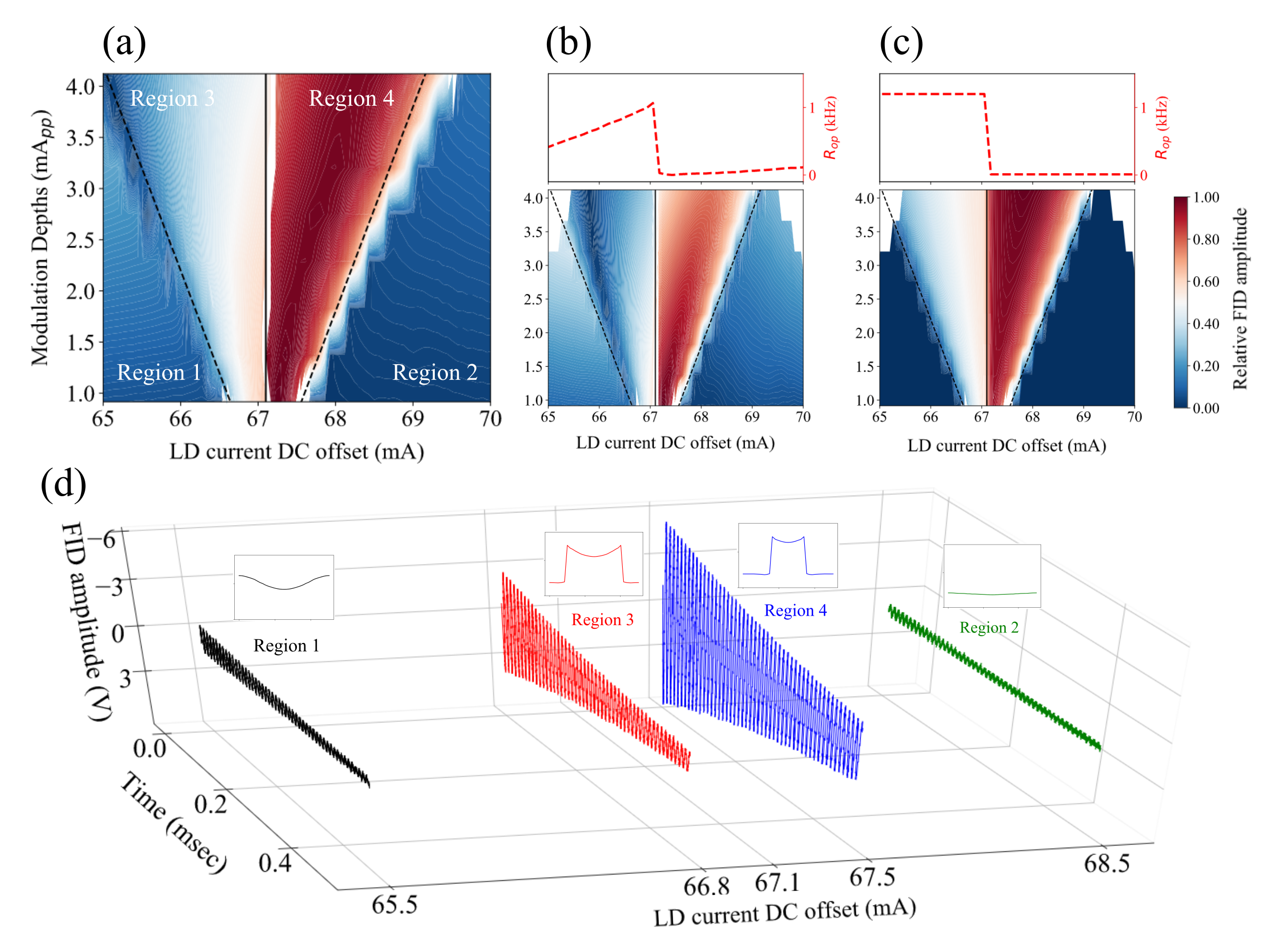}
\caption{(a) FID amplitude as a function of LD current DC offsets and modulation depths. The results are divided into four regions by two dashed lines (the boundaries where modulation includes a mode-hop current) and a straight line (the mode-hop current at 67.1 mA). (b) A semi-theoretical result. $S_0$ is calculated based on the optical pumping profile shown in the top panel. The optical pumping rate profile is extracted by subtracting the minimum relaxation rate from the measured relaxation rate shown in Fig.~\ref{fig:absorption}~(c). (c) The semi-theoretical result is based on the step-function-like optical pumping profile. (d) FID signals at four regions. The insets show 1-period temporal pumping rate profiles extracted from the top panel of (b). The modulation depth is fixed at 1.8 mA$_{pp}$.}
\label{fig:modehopeff}
\end{figure*}

\section{Experimental results and discussion}

 Changes in the LD injection current lead to concurrent changes in laser beam intensity and detuning, consequently affecting both the Faraday rotation angle and the spin relaxation rate. To examine these parameters as a function of the LD injection currents, we obtained FID signals by changing the LD current DC offset at the probe stage, while current modulation at the pumping stage remained fixed. Because the spin polarization is unchanged (due to fixed current modulation at the pumping stage), the FID signal reflects the atomic response on the LD current DC offset. Fig.~\ref{fig:absorption}~(a) depicts the absorption spectra of ${}^{87}$Rb in the atomic vapor cell. Fig.~\ref{fig:absorption}~(b) shows the laser frequency detuning as a function of LD injection currents. A huge shift in the laser frequency occurs at around 67.1 mA. 
 
 The relative size of Faraday rotation angles and transverse spin relaxation rates as a function of LD injection currents is shown in Fig.~\ref{fig:absorption}~(c). At a current lower than the mode-hop current, the laser  becomes nearly resonant so that a large Faraday rotation angle and a fast spin relaxation are induced. Conversely, with an LD current higher than the mode-hop current, the laser frequency shifts into off-resonance, leading to a relatively small Faraday rotation angle and a smaller spin relaxation rate. The maximum difference in spin relaxation rates between on- and off-resonance was approximately 1036 /s, which represents the maximum optical pumping rate. The minimum spin relaxation rate in off-resonance is 2392~ /s when the $1/e^2$ beam diameter is 10 mm, which is the internal size of the atomic vapor cell. When the beam diameter is small, Rb-Rb spin exchange collisions act as the main spin relaxation mechanism. A polarized atomic spin in the boundary between the bright region and the dark region is easily depolarized by a spin exchange collision with an unpolarized atomic spin from the dark region. With a beam diameter of 1.5 mm, the minimum spin relaxation rate was measured as 4429~ /s which is very close to 4304~ /s, the \(^{87}\)Rb spin-exchange relaxation rate at the cell temperature of 98 ${}^\circ$C \cite{gibbs1967spin}. This means that the optical pumping can be effectively turned off by using laser mode hopping. Fig.~\ref{fig:absorption}~(c) shows the optical power of the laser beam after passing through the atomic vapor cell. The absorption is minimized in the off-resonance condition (laser current above the mode-hop current), leading to a higher FID amplitude $V_0$ despite the slight degradation in the Faraday rotation angle.

We measured the initial amplitude of the FID as a function of the LD current DC offset and the current modulation depth as shown in Figure.~\ref{fig:modehopeff}. Fig.~\ref{fig:modehopeff}~(a) shows the measured relative FID amplitude. We divide Fig.~\ref{fig:modehopeff}~(a) into the four regions, depending on whether the current modulation in the pumping stage includes the laser mode-hop current and whether the current in the probing stage is larger (or smaller) than the laser mode-hop current. The black dashed lines in Fig.~\ref{fig:modehopeff}~(a)-(c) indicate the boundary where the current modulation in the pumping stage contains the laser mode-hop current. The black solid line depicts the mode-hop current, 67.1 mA. The laser frequency in the probing stage is off-resonant (or on-resonant) in the region to the right (or left) of the black solid line. 
We note that the experimental data shown in Fig. 4 (a) reflects the combined effect of the paramagnetic Faraday rotation, beam absorption and the discontinuous optical pumping.

In region 1, the current modulation in the pumping stage excludes the laser mode-hop current so that the continuous optical pumping profile with a small $R_1$ causes a small $S_0$. Additionally, the current in the probing stage brings the laser frequency into on-resonance, resulting in a small beam intensity due the strong absorption. These factors lead to the small FID amplitude as depicted by the black line in Fig.~\ref{fig:modehopeff}~(d). 
In region 2, the current modulation in the pumping stage does not contain the mode-hop current, as in region 1. However, the value of $R_1$ in region 2 is smaller than that in region 1 as shown in the insets of Fig.~\ref{fig:modehopeff}~(d). Hence, $S_0$ of the region 2 is expected to be smaller than that of region 1. On the other hand, the laser frequency in the probing stage is off-resonant, resulting in a spin relaxation rate that is much smaller than that in region 1. The resulting FID signal is depicted by the green line in Fig.~\ref{fig:modehopeff}~(d), whose amplitude is slightly smaller than region 1 (black line). 
In region 3, the mode-hop current is included in the current modulation, resulting in a highly discontinuous modulation profile. A large $S_0$ is produced and a large Faraday rotation occurs in the probing stage. In this region  the laser frequency in the probing stage is on-resonance. The absorption of the laser beam, therefore, results in a degradation in FID amplitude, as shown by the red line in Fig.~\ref{fig:modehopeff}~(d).
The maximum FID amplitude is in region 4, as shown by the blue line in Fig.~\ref{fig:modehopeff}~(d). A large $S_0$ is induced because the current modulation, including the laser mode-hop current, results in a discontinuous optical pumping profile with a large $R_1$. The off-resonant beam minimizes absorption in the probing stage, inducing a large FID amplitude compared to region 3. In addition, the spin relaxation rate in region 4 is slower than in region 3, which is beneficial for magnetometer sensitivity.

To compare the experimental data with the theory, Eq.~\eqref{SpinAmplitude} is calculated using \(R_0\) and \(R_1\) extracted from the \(R(t)\) profile shown at the top panel of Fig.~\ref{fig:modehopeff}~(b) and (c). The \(R(t)\) profile in the top panel of Fig.~\ref{fig:modehopeff}~(b) is obtained by subtracting the minimum relaxation rate from the measured spin relaxation rate shown in Fig.~\ref{fig:absorption}~(c). Fig.~\ref{fig:modehopeff}~(b) shows the calculated relative FID amplitude using the triangular function like \(R(t)\) profile in the top panel of Fig.~\ref{fig:modehopeff}~(b). Fig.~\ref{fig:modehopeff}~(c) is derived from the fictitious step-function like \(R(t)\) profile given in the top panel of Fig. 4 (c). Both the experimental result and the semi-theoretical results seem to have similar tendencies. But the details are slightly different from each other. Figure 5 shows the maximum FID amplitudes as a function of the modulation depth and the LD current DC offsets with the maximum FID amplitude in the experiment (Fig.~\ref{fig:modehopeff}~(a)) and two semi-theoretical calculations (Fig.~\ref{fig:modehopeff}~(b) and (c) ). Since the two semi-theoretical results and the experimental result use the same Faraday rotation profile and the same laser beam absorption profile, the differences between them are mainly attributed to the discontinuous optical pumping.

\begin{figure}
\centering
\includegraphics[width=0.9\linewidth]{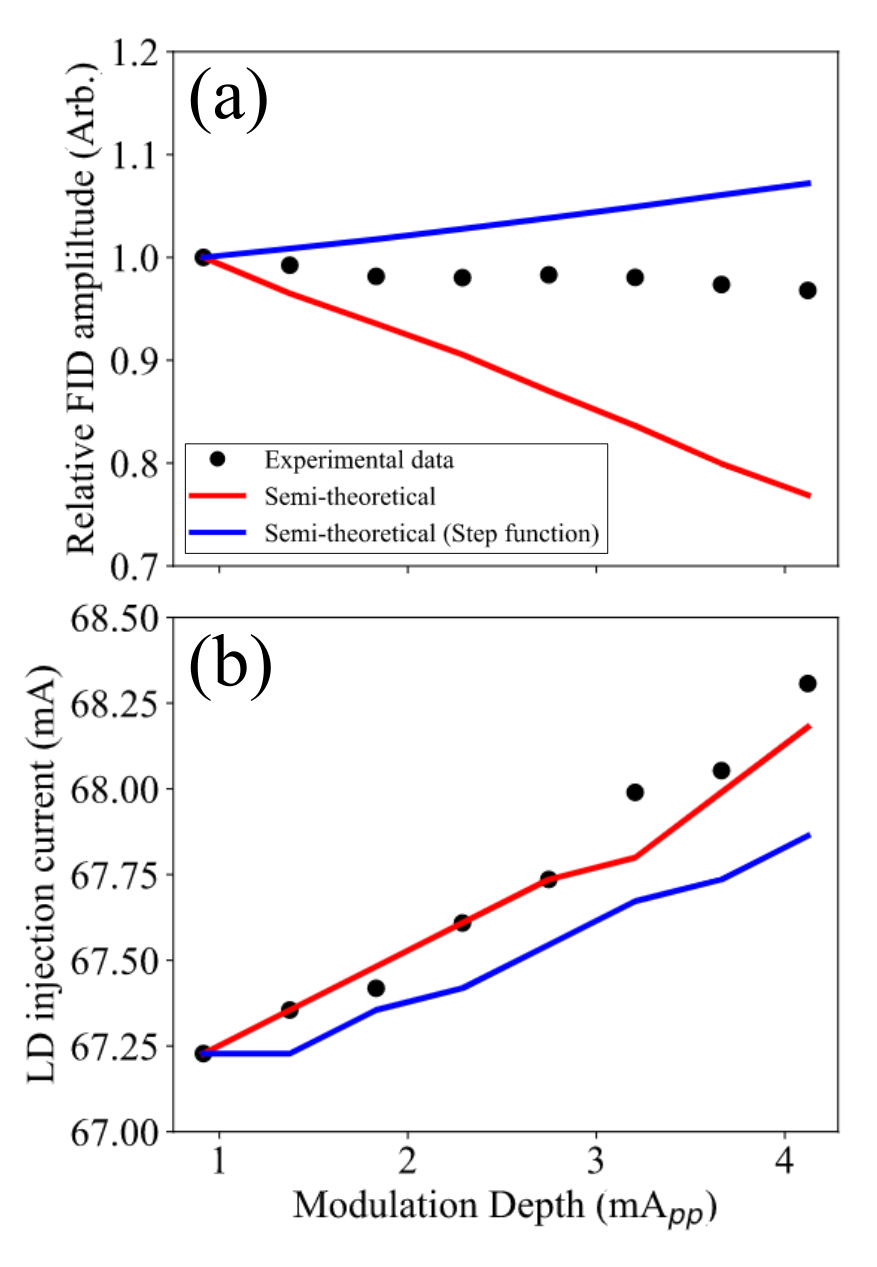}
\caption{(a) Maximum FID amplitude as a function of modulation depth. The black dots represent the experimental data, derived from Fig.~\ref{fig:modehopeff}~(a). The red line and the blue line depict the semi-theoretical results extracted from Fig.~\ref{fig:modehopeff}~(b) and Fig.~\ref{fig:modehopeff}~(c), respectively. The experimental data (black dots) is more robust to the change in modulation depth compared to the semi-theoretical result (red line) where the optical pumping profile is a triangular function. The semi-theoretical result (blue line) where the optical pumping profile is a step function shows that the maximum amplitude of FID increases with increasing modulation depth. (b) LD current DC offset with the maximum FID amplitude as a function of modulation depth.}
\label{fig:peak}
\end{figure}

Comparing the experimental data (black dots) and semi-theoretical result with the measured triangular function like optical pumping profile (red line) in Fig. 5 (a), the maximum FID amplitude decreases moderately as the modulation depth grows while $S_0$ derived by Eq.~\eqref{SpinAmplitude} rapidly decreases. This implies that the measured triangular function like optical pumping profile used in the top panel of Fig.~\ref{fig:modehopeff} (b) does not fully describe the mode-hopping assisted optical pumping. 

This discrepancy between the semi-theoretical result and the experimental result needs to be studied in the future. Here, we describe its potential origins. First, we note that the optical pumping rate profile shown in Fig.~\ref{fig:modehopeff} (b) is calculated from the transverse spin relaxation rates extracted from the FID signals as a function of the LD current DC offset in the probing stage. The FID signals are affected by the constant laser frequency. However, because the laser frequency sweeps in the pumping stage, the optical pumping rate in the pumping stage is slightly different from the optical pumping rate derived from the FID signals as a function of the LD current DC offset in the probing stage. In particular, when the ground hyperfine structure is considered, the optical pumping rates of the $F_g = 1$ state and the $F_g = 2$ change with times because of the laser frequency sweeping. The transverse spin along the \(\hat{x}\)-direction can be represented by $\langle S_x \rangle = ( \langle F_x^{I+1/2} \rangle - \langle F_x^{I-1/2} \rangle )/(2I+1)$ where $I$ represents the nuclear spin, $I = 3/2$ for $^{87}$Rb. To maximize spin polarization $S_0$ of $^{87}$Rb,  the $F_g = 2, ~m_{F_g} = 2$ state (or $m_{F_g} = -2$) should be mainly populated. For this, the hyperfine repumping of the $F_g = 1$ state is required~\cite{li2022repumping, schultze2015improving}. Due to the large pressure broadening on the order of GHz, the laser at the pumping stage pumps both the $F_g = 1$ and $F_g = 2$ states with different optical pumping rates. The different time-varying optical pumping rates for the $F_g = 1$ and the $F_g = 2$ states have to be taken into account to describe the mode-hopping assisted optical pumping more accurately. 

Second, the spatial inhomogeneity of the optical pumping profile needs to be considered. Due to light absorption, the optical pumping beam is attenuated along the beam propagation direction so that the optical pumping profile is spatially inhomogeneous. The measured optical pumping profile shown in the top panel of Fig.~\ref{fig:modehopeff} (b) is a kind of spatially averaged values. The analysis considering the spatially inhomogeneous optical pumping profile will give more accurate results. Additionally, we obtained the semi-theoretical result with the step-function-like optical pumping profile, as depicted by the blue solid line in Fig.~\ref{fig:peak}~(a). In contrast to the experimental result and the semi-theoretical result with the triangular function like optical pumping profile, the relative FID amplitude increases with increasing modulation depth. 

Interestingly, the experimental data lie between two semi-theoretical results. We can conjecture that the real optical pumping profile lies between the triangular function and the step function. 

\begin{figure*}
\centering
\includegraphics[width=0.99\linewidth]{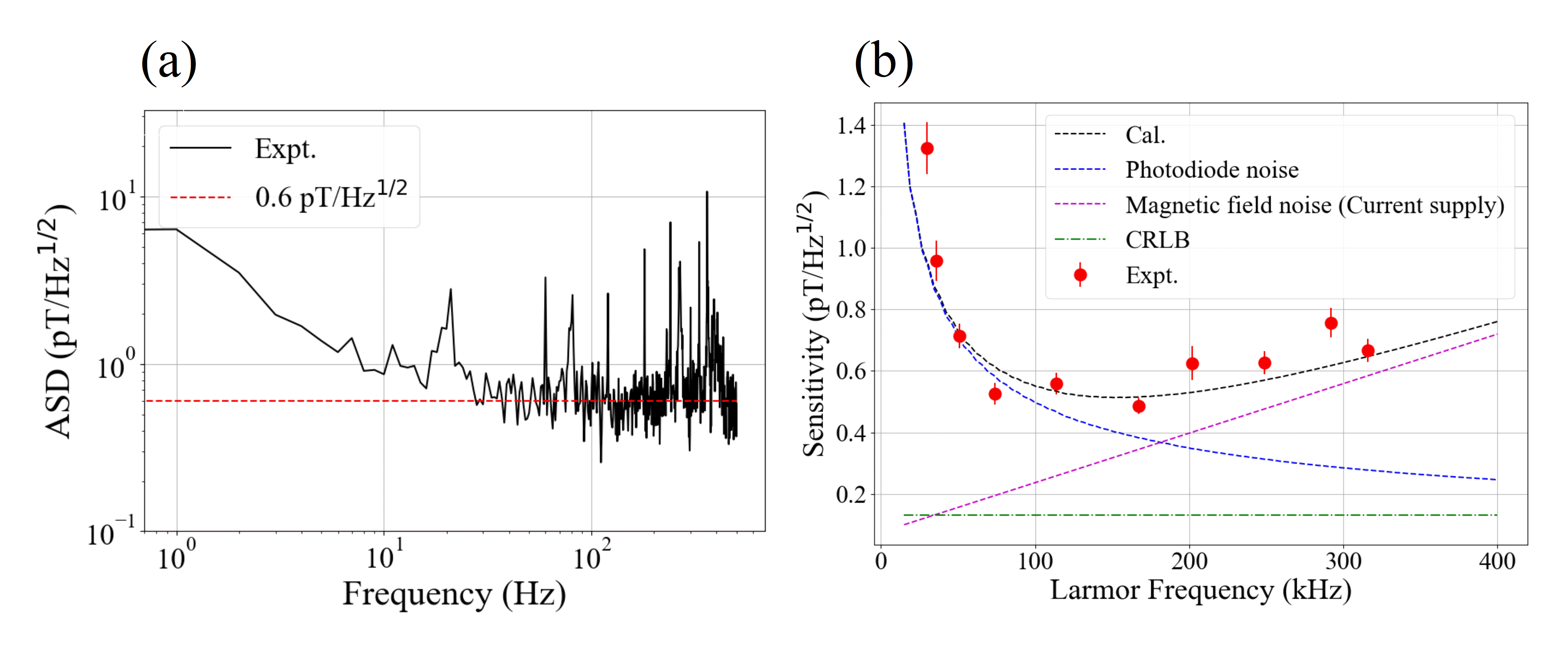}
\caption{(a) Amplitude spectral density of the magnetic field noise at a bias field of 27 $\mu$T. The red dashed line represents a magnetic noise floor of 0.6 pT/Hz$^{1/2}$. (b) Noise as a function of the Larmor frequencies. The blue dashed line depicts the calculation for frequency counting using Eq.~\eqref{eq:sensitivity}. The magenta dashed line represents the magnetic field noise generating from the bias magnetic field coil driven by the current supply with a current noise of 11.3 nA/Hz$^{1/2}$/A. The green dashed-dot lines shows the Cram\'{e}r-Rao lower bound (Eq.~\eqref{CRLBEq}), 0.131 pT/Hz$^{1/2}$  in our experimental conditions. }
\label{fig:sensitivity}
\end{figure*}

We estimate the sensitivity of our magnetometer based on the laser mode-hopping assisted optical pumping. In order to further decrease the effect of the unwanted external magnetic field and to increase the uniformity of the magnetic field, we applied the 4 layered \(\mu\)-metal shield (MS-1, Twinleaf) instead of the small home-made 2 layered \(\mu\)-metal shield which were used to test the functionality of the laser mode-hopping. After applying the bias magnetic field along the z-axis of the magnetic shield, we measured Larmor frequencies at a repetition rate of 1 kHz by using the commercial frequency counter. The measured Larmor frequencies were converted into the magnitude of the magnetic field by diving the Larmor frequencies by the gyromagnetic ratio of $^{87}$Rb. Finally, the amplitude spectral density (ASD) of the magnitude of the magnetic field was calculated as shown in Fig.~\ref{fig:sensitivity} (a). The magnetic field noise floor level is 0.6 pT/Hz$^{1/2}$ at a magnetic field of 27 $\mu$T, as depicted by the red dashed line in Fig.~\ref{fig:sensitivity} (a). 

In order to observe the magnetic field noise floor limited by the frequency counting method based on the linear regression of zero-crossing times, we obtained the ASDs as a function of Larmor frequencies. The Larmor frequency changes depending on the current flowing in the bias coil. When the frequency is extracted from the linear regression of zero-crossing times, the magnetic field noise can be calculated as 
\begin{equation}
\rho_{fc} = \frac{12 \delta y ~f_0}{2 \pi \gamma V_0  N^{3} \sqrt{ f_{BW}} } \sqrt{\sum_{j}^{N} {(j-\frac{\sum j}{N})^2} e^{2 j \Gamma_2 /f_0}},
\label{eq:sensitivity}
\end{equation}
where $j$s are integer numbers and $N$ is the total number of the zero-crossings, given by $N = f_0  T_m$ where $f_0$ is the Larmor frequency and $T_m$ is the measurement time. The total rms amplitude noise $\delta y$ is given  by $\delta y = \sqrt{\delta^2 y_{fc}+\delta^2 y_{ph}}$ where $\delta y_{fc}$ is the amplitude noise of the frequency counter, 350 $\mu$$\textrm{V}_{\textrm{rms}}$ and $\delta y_{ph}$ is the RMS photon-shot-noise. The amplitude spectral density of the photon-shot-noise is given by  $\sqrt{2eG^2RP_0}$, where $G$ is the transimpedence gain ($G=10^{5}$ V/A for a high-Z load), $R$ is the responsivity of the BPD, $P_0$ is the optical power in front of the polarimeter, and $e$ is the electron charge. When the total power of the probe beam after passing through the atomic gas cell was 4 mW, the measured ASD including the probe electronic noise and the photon-shot-noise was 2.0 $\mu$V/Hz$^{1/2}$.  It gave a RMS voltage of 1400 $\mu$$\textrm{V}_{\textrm{rms}}$ in the fixed noise band width of $N_{BW} = 500$ kHz. The magnetic field noise originating from the electronic noise and the photon-shot-noise, as a result of Eq.~\eqref{eq:sensitivity},  is represented by the blue dashed line in Fig.~\ref{fig:sensitivity} (b). In opposition to the result of Eq.~\eqref{eq:sensitivity},the measured noise floor increases with the averaged current. 

In order to investigate such linear dependence of the magnetic field noise on the averaged current, we considered the magnetic field noise driven by the current noise of the current supply (CSUA-300, Twinleaf). We measured the current noise of the current supply by using a signal analyzer (N9010A, Keysight). The current noise was linearly proportional to the averaged current $\left< I \right>$ and it can be represented as $\delta I = \alpha + \beta \left<I\right>$ where $\alpha = 0.6$ (0.08) nA/Hz$^{1/2}$ and $\beta = 11.3$ (0.8) nA/Hz$^{1/2}$/A. By considering the coil calibration constant, 130 $\mu$T/A, the magnetic field noise due to the current noise of the current supply can be estimated by $\delta B_{cc} =$  1.47 pT/Hz$^{1/2}$/A $\times \left< I \right> $ + 0.078 pT/Hz$^{1/2}$ . The magenta dashed line in Fig.~\ref{fig:sensitivity} (b) depicts the magnetic field noise due to the current noise from the current supply. The squared sum combination of the photon-shot-noise and the magnetic field noise from the current noise of the current supply, $\delta B = \sqrt{\rho^2_{fc} + \delta B_{cc}^2 }$, is represented by the black dashed line in Fig.~\ref{fig:sensitivity} (b). It shows a good agreement with the experimentally measured magnetic field noise floor. 

Our frequency measurement using the commercial frequency counter was not optimal. The potentially achievable sensitivity can be estimated by the CRLB, calculated using a shot of the FID signal \cite{gemmel2010ultra}. The CRLB is the sensitivity reference for an atomic magnetometer based on measurement of the Larmor frequency of the decaying FID signal. The CRLB is given by 

\begin{equation}
\rho_{CRLB} = \frac{\sqrt{12 C_w} \rho_\omega}{\gamma V_0 \sqrt{f_{BW}} T_m^{3/2}}.
\label{CRLBEq}
\end{equation}

\noindent
where $\rho_\omega$ is the ASD level of the BPD signal at the Larmor frequency. The coefficient $C_\omega$ which depends on the transversal spin relaxation rate $\Gamma_{re}$ which is given by

\begin{equation}
C_w = \frac{e^{2/r}-1}{3r^3 \textrm{cosh}(2/r) - 3r(r^2+2)}.
\end{equation}
where $r = 1/(\Gamma_{re} T_m)$. The CRLB calculated by Eq.~\eqref{CRLBEq} represents the non-magnetic oriented noise induced sensitivity.  After optimization, the parameters for calculating the CRLB are given as $\Gamma_{re} = 1670$ /s, $T_m = 650$ $\mu$s, $\rho_{\omega} = 2 $ $\mu$V/Hz$^{1/2}$, $V_0 = 8$ V, and $f_{BW} = 500$ Hz in our experiments using the four layered \(\mu\) shield. We obtain the CRLB of 131 fT/Hz$^{1/2}$. The CRLB calculation shows that our magnetometer is potentially capable of reaching a sensitivity of a sub pT/Hz$^{1/2}$ by revising the frequency counting method; the noise bandwidth is adjusted to the Larmor frequency, { \it{i.e.}}, $N_{BW} \simeq f_0$. 

We note that the frequency shift of 20 GHz, and the laser frequency in the pumping stage which is moderately detuned from the resonance, are not optimal. In order to mainly populate the $F_g = 2,~ m_{F_g} = 2$ state to maximize $S_0$, the on-resonance laser frequency in the pumping stage should be close to the $F_g = 1, 2$ resonances. Also the laser frequency needs to be periodically far-detuned from the resonance for synchronous optical pumping, to obtain a large optical pumping modulation depth. A careful laser cavity adjustment in the laser diode fabrication, leading to an optimal frequency shift of mode-hopping and the on-resonance laser frequency very close to the resonance, will be helpful to increase the sensitivity of the magnetometer. In addition, multi-pass probing will further increase the sensitivity because $V_0$ is proportional to the interaction length \cite{li2011optical}.

\section{Conclusion}
We have demonstrated  a single-beam all-optical pulsed atomic magnetometer enhanced by laser mode-hopping. When optical pumping is assisted by laser mode-hopping, it leads to a discontinuous pumping profile, effectively enhancing the transverse spin polarization. Furthermore, mode-hopping offers a convenient method for setting  the probing stage beam off-resonant, which suppresses absorption and maintains a large FID signal. We achieved a sensitivity of 0.6 pT/Hz$^{1/2}$ which is sufficient for applications to detect magnetic anomalies in earth's magnetic field. The sensitivity of our magnetometer seems to be limited by photon-shot-noise and the magnetic field noise induced by the current noise in the current supply for driving the bias magnetic field coil.  In order to estimate the potentially achievable sensitivity, the CRLB due to the non-magnetic oriented noise was calculated, which was 131 \(\textrm{fT}/\sqrt{\textrm{Hz}}\) in our experimental condition. Considering the calculated CRLB, we expect that our magnetometer can reach a sensitivity of a sub pT/Hz$^{1/2}$ which is similar to sensitivity limited by photon shot-noise.  Laser cavity optimization considering laser mode-hopping during the fabrication of a DBR LD may further improve sensitivity by enhancing optical pumping. Our approach based on laser mode-hopping can be applied for  the miniaturization of all-optical atomic magnetometers with sub-pT/Hz$^{1/2}$ sensitivities.

\begin{section}{Acknowledgments}
This work was supported by the Agency for Defense Development Grant funded by the Korean government (No. 915098102).
\end{section}

\section{Appendix}

\subsection{Consideration of Faraday Rotation of Elliptically Polarized Beam}

The spin polarization formed assisted by mode-hopping enters the probing stage where elliptically polarized light undergoes optical Faraday rotation, then measured by a polarimetry. With a linear polarization basis ($\hat{e}_H, \hat{e}_V$), the elliptical polarization state is represented as \((\cos(\theta), \sin(\theta) \exp(i \phi))\). Here, \(\theta\) is the tilt angle, and \(\phi\) is the retardation angle. The change in polarization state of the beam after passing through the atomic vapor can be expressed by the matrix 
\(M_c = \begin{pmatrix}{1+e^{i \Phi}}&{i(1-e^{i \Phi})}\\{i(e^{i \Phi}-1)}&{1+e^{i \Phi}}\end{pmatrix} \) 

where 
\(\Phi = |s|\Phi_0(t)\) is the Faraday rotation angle. Here, \(\Phi_0(t)\) is damped oscillation in our experiment and \(|s|\) is the size of the photon spin vector which is calculated by \(\sin(2\theta)\sin(\phi)\). The matrix representation of the HWP is \(M_h = \begin{pmatrix}{\cos (2\theta_h) }&{\sin(2\theta_h) }\\{\sin(2\theta_h) }&{-\cos(2\theta_h)}\end{pmatrix}\), so the final polarization state of the elliptically polarized beam measured by the polarimeter is \(M_h \cdot M_c \cdot (\cos (\theta), \sin(\theta)\exp(i \phi))\). The balanced signal is then calculated by the difference between the squares of each orthogonal linear polarization components of the beam: \(4 I_0 [\cos(4\theta_h+\Phi)\cos(2\theta) +\cos(\phi)\sin(4\theta_h+\Phi) \sin(2\theta)]\). Here, \(I_0\) is the intensity of the beam after absorption throughout the vapor cell. When \(1 \gg \Phi\), the oscillating signal simplifies to
\begin{multline}
\label{QWPeq}
I_B \approx I_0[
\cos(4\theta_h)\cos(2\theta) +\sin(4\theta_h)\sin(2\theta)\cos\phi \\
+|s| \Phi_0(t)(
-\sin(4\theta_h)\cos(2\theta)+\cos(4\theta_h)\sin(2\theta)\cos\phi) ].
\end{multline}

\begin{figure}[t]
\centering
\includegraphics[width=0.9\linewidth]{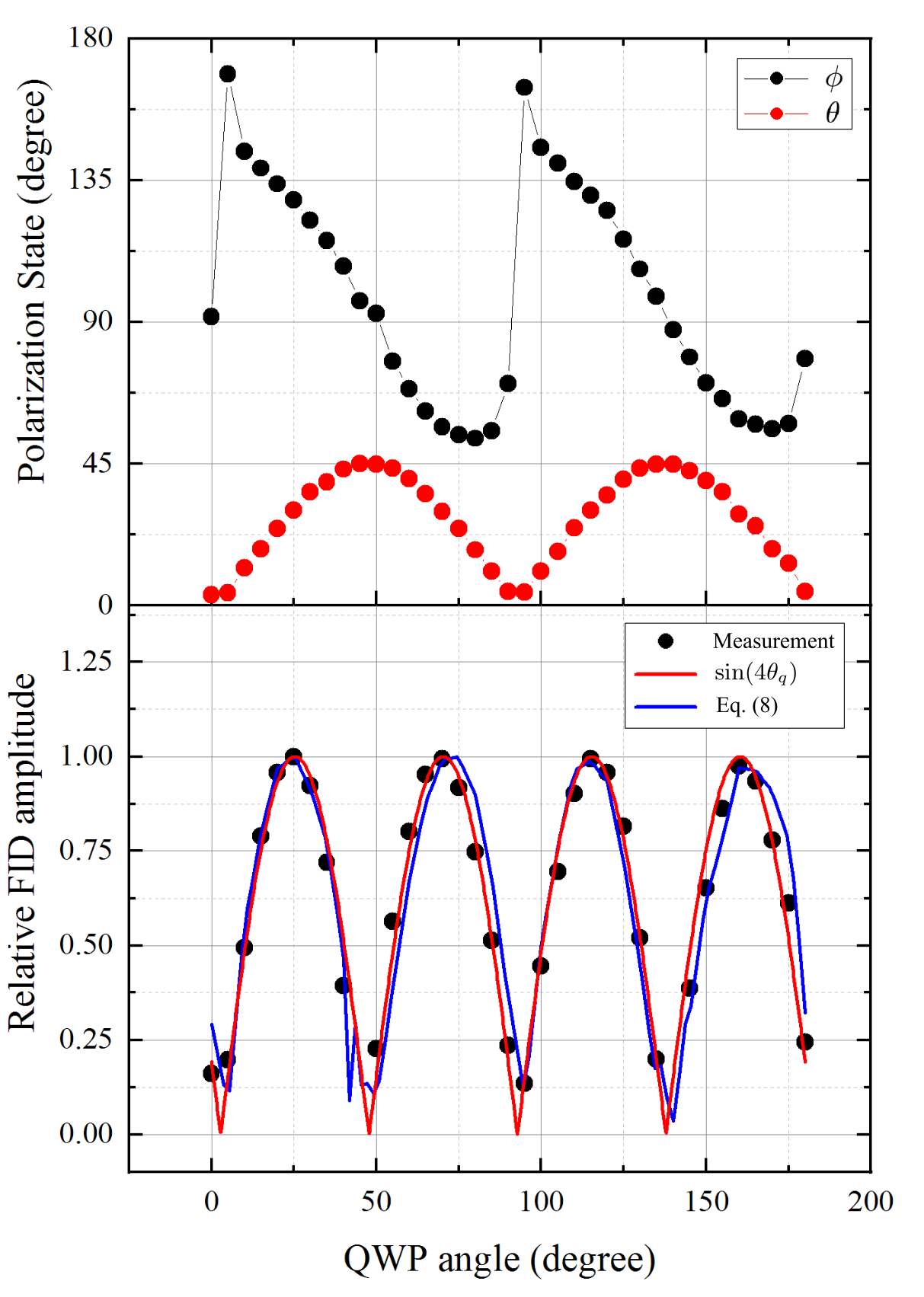}
\caption{The polarization state of  the incident beam (upper plot) and relative FID amplitude (bottom plot) as a function of QWP angle. When the QWP angle is 25 ${}^\circ$ where the relative FID amplitude is maximized, $\theta$ is 30 ${}^\circ$ and $\phi$ is 129 ${}^\circ$. The red line shows $\sin(4\theta_q)$ behavior to the QWP angle $\theta_q$. The blue line illustrates Eq.~\eqref{QWPeq} calculated based on the polarization state on the top plot.}
\label{fig:QWPangle}
\end{figure}

\noindent
By adjusting the HWP angle \(\theta_h\), the baseline component of the balanced signal can be eliminated. The amplitude is coefficient proportional to \(\Phi_0(t)\), and is maximized at the same time. 

A previous study has shown that signal amplitude in single-beam SERF magnetometers follows a \(\sin(4\theta_q)\)-dependence on the QWP angle \(\theta_q\)\cite{shah2009spin}. Our all-optical single beam pulsed magnetometer shows similar behavior under the bias field of 14.3 \(\mu T\) which is comparable to the Earth's ambient magnetic field. The top plot of Figure.~\ref{fig:QWPangle} shows the polarization state of the beam before entering the vapor cell as measured by a polarimeter (PAX 5710IR1, Thorlabs) at various QWP angles. The bottom plot shows the relative FID amplitude measured at different QWP angles. The red line illustrates \(|\sin(4\theta_q)|\) behavior, while the blue line represents equation~\eqref{QWPeq}, which is calculated based on the polarization state shown in the upper plot. The HWP angle \(\theta_h\) is adjusted to compensate so the signal baseline will be zero for each measurement.

We determined the optimal QWP angle for the single-beam all-optical magnetometer to be 25 ${}^\circ$ with a retardation of 133 ${}^\circ$, suggesting an elliptically polarized beam as the most effective polarization state. The elliptical polarization strikes a balance between the most efficient pumping of circular polarization, and the most effective Faraday rotation that occurs with linear polarization. Note that baseline compensation by adjusting the HWP angle is essential for FID measurements. Without baseline compensation, the FID amplitude may decrease, and if the beam intensity is too high, the baseline component may saturate the entire signal. To avoid signal saturation on the BPD signal, in our measurement, the beam intensity in front of the vapor cell is kept at 1.43 mW during this measurement.

\bibliographystyle{unsrt}
\section{References}
\bibliography{ref}

\end{document}